\documentclass[twoside,a4paper,11pt]{sca}
\usepackage{graphicx}
\usepackage{hyperref}
\usepackage{movie15}
\usepackage{natbib}  % Cross-reference package (Natural BiB)
\begin{document}
\pagenumbering{arabic}
\pagestyle{myheadings}
\thispagestyle{empty}
%%%%{\flushright\includegraphics[width=\textwidth,bb=58 650 590 680]{stamp.pdf}}
{\flushright\includegraphics[width=\textwidth,bb=90 650 520 700]{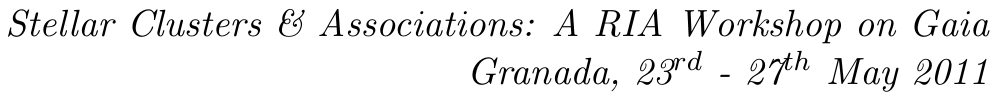}}
\vspace*{0.2cm}
\begin{flushleft}
{\bf {\LARGE
%
%%% TITLE of the paper. 
Stellar Clusters in M31 from PHAT: Survey Overview and First Results
%
% Do not delete next few lines
}\\
\vspace*{1cm}
%
%%% LIST OF AUTHORS.
L. Clifton Johnson$^{1}$, Anil C. Seth$^{2}$, Julianne J. Dalcanton$^{1}$, Nelson Caldwell$^{2}$, Dimitrios A. Gouliermis$^{3}$, Paul W. Hodge$^{1}$, S{\o}ren S. Larsen$^{4}$, Knut A. G. Olsen$^{5}$, Izaskun San Roman$^{6}$, Ata Sarajedini$^{6}$, Daniel R. Weisz$^{1}$, and the PHAT Collaboration
%
% Do not delete next few lines
}\\
\vspace*{0.5cm}
%
%%% AFFILIATIONS LIST.
$^{1}$
Department of Astronomy, University of Washington, Box 351580, Seattle, WA 98195, USA\\
$^{2}$
Harvard-Smithsonian Center for Astrophysics, 60 Garden Street Cambridge, MA 02138, USA\\
$^{3}$
Max Planck Institute for Astronomy, K\"onigstuhl 17, 69117 Heidelberg, Germany\\
$^{4}$
Astronomical Institute, University of Utrecht, Princetonplein 5, NL-3584 CC, Utrecht, The Netherlands\\
$^{5}$
National Optical Astronomy Observatory, 950 North Cherry Ave., Tucson, AZ 85719, USA\\
$^{6}$
Department of Astronomy, University of Florida, 211 Bryant Space Science Center, Gainesville, FL 32611-2055, USA

%
% Do not delete next few lines
\end{flushleft}
%
% Headings
\markboth{
%%% Type the SHORT version of the paper title.
Stellar Clusters in M31 from PHAT
}{ % Do not delete
%
%%%  First Author \& Second Author   OR   First-author et al. if the author list contains three or more authors.
Johnson et al.
% 
% Do not delete next few lines
}
\thispagestyle{empty}
\vspace*{0.4cm}
\begin{minipage}[l]{0.09\textwidth}
\ 
\end{minipage}
\begin{minipage}[r]{0.9\textwidth}
\vspace{1cm}
\section*{Abstract}{\small
%
%%% ABSTRACT
The Panchromatic Hubble Andromeda Treasury (PHAT) is an on-going Hubble Space Telescope (HST) multi-cycle program that will image one-third of the M31 disk at high resolution, with wavelength coverage from the ultraviolet through the near-infrared.  This dataset will allow for the construction of the most complete catalog of stellar clusters obtained for a spiral galaxy.  Here, we provide an overview of the PHAT survey, a progress report on the status of observations and analysis, and preliminary results from the PHAT cluster program.  Although only $\sim$20\% of the survey is complete, the superior resolution of HST has allowed us to identify hundreds of new intermediate and low mass clusters.  As a result, the size of the cluster sample within the Year 1 survey footprint has grown by a factor of three relative to previous catalogs.
%
% Do not delete next few lines
\normalsize}
\end{minipage}
%
%
%%% BODY

\section{Introduction \label{intro}}
The Andromeda galaxy (M31) is an exquisite laboratory for studying stellar clusters.  Its proximity \citep[785 kpc;][]{McConnachie05} allows for the detailed study of individual stars in clusters, while simultaneously providing a large, galaxy-wide sample of objects.  Studies of M31's stellar cluster system have been on-going since the work of \citet{Hubble32}.  The Panchromatic Hubble Andromeda Treasury (PHAT) survey, described in Sec.~\ref{phat}, represents a significant step forward in the study of Andromeda's cluster system, extending the sample of known clusters well into the intermediate and low mass regimes.  This dataset will inform our understanding of cluster evolution as a whole, through its wide sampling of age, mass, and galactic environment parameter space.  The survey's cluster science goals include placing constraints on cluster disruption behavior, assessing environmental dependencies of cluster formation and destruction, and measuring the star cluster initial mass function, among many others.  The first step to achieving these goals lies in the accurate identification and characterization of M31's cluster system.  We describe our current progress on this task in Sec.~\ref{clusterwork}, and direct the reader to our forthcoming paper (Johnson et al., in prep.) for complete details.

\section{The PHAT Survey\label{phat}}
\begin{figure}
\center
\includegraphics[scale=0.42]{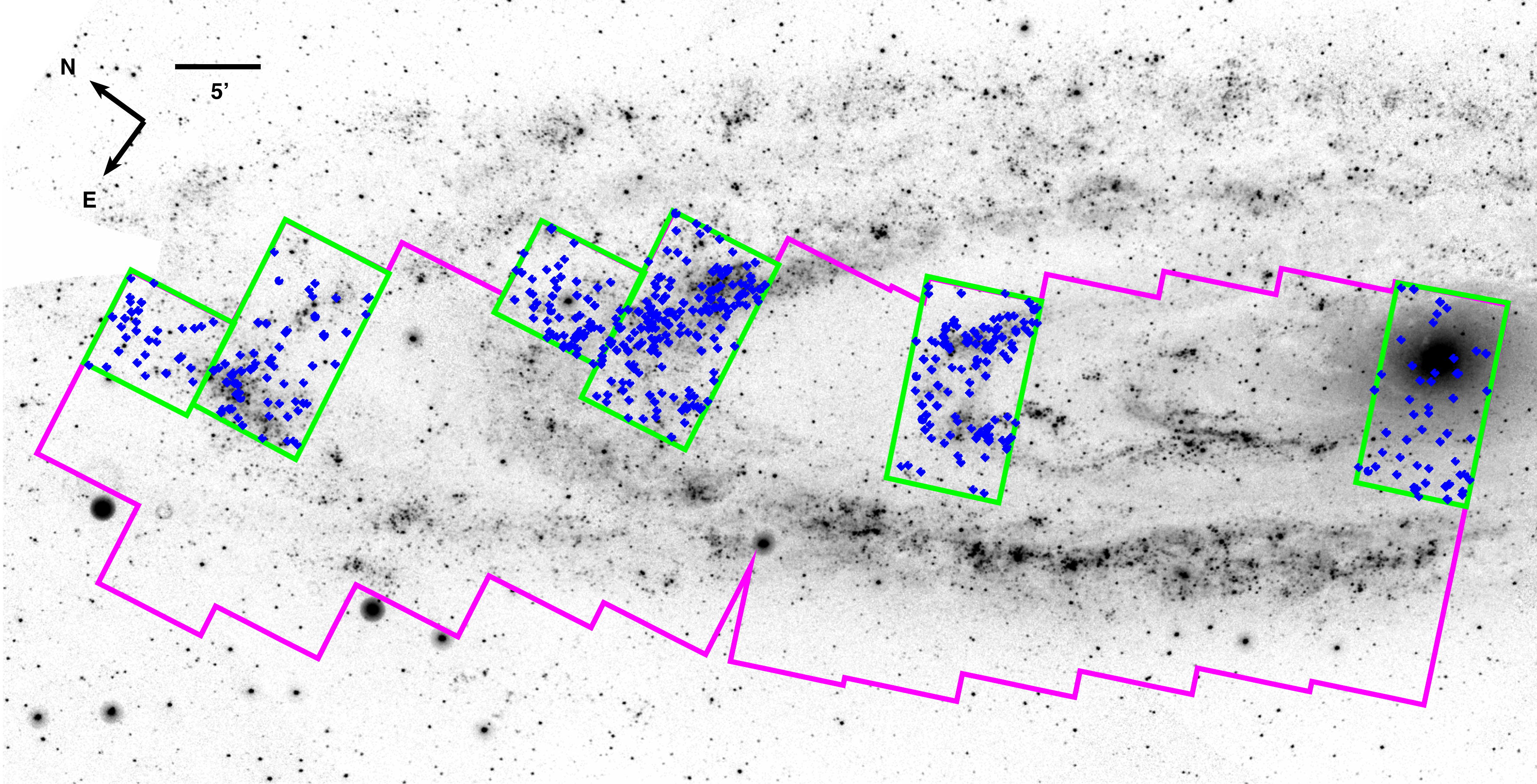} 
\caption{\label{footprint} The footprint of the PHAT survey region (magenta) displayed on a GALEX NUV image of the northeast half of M31.  Green rectangles represent the ``bricks" that make up the Year 1 imaging data.  Blue circles show the spatial distribution of cluster identifications resulting from our Year 1 by-eye search.}
\end{figure}

The PHAT survey\footnote{Project Website: \href{http://www.astro.washington.edu/groups/phat}{http://www.astro.washington.edu/groups/phat}} (PI: Dalcanton) is a Hubble Space Telescope (HST) multi-cycle treasury program that will image one-third of M31's stellar disk.  The survey region extends across the northeast half of M31, resulting in areal coverage stretching from the galactic nucleus out to the edge of the star-forming disk at galactocentric radii of $\sim$20 kpc.  The full survey footprint is shown in Fig.~\ref{footprint}.  Imaging across $\sim$0.5 deg$^2$ of contiguous spatial coverage will be obtained using three different HST instruments (WFC3-UVIS, ACS-WFC, WFC3-IR) in six filters ranging from the ultraviolet to the near-infrared (UV to NIR; F275W, F336W, F475W, F814W, F110W, F160W).  Observations are grouped into 23 units known as ``bricks", each of which is made up of 18 tiled fields-of-view arranged in two side-by-side, 3$\times$3 half-brick arrays.  Bricks are observed over two epochs, separated by six months, in which the optical imaging is collected for one of the half-brick arrays, and the UV and NIR images in the other.  Six months later, these wavelength assignments are reversed in accordance with a 180-degree difference in HST roll angle, and six-filter imaging coverage for the brick is completed.  For a complete survey and data reduction description, please see Dalcanton et al. (in prep.).

The PHAT survey has been allocated more than 800 HST orbits, which will be executed over the course of four years.  As of May 2011, four full bricks and two additional half-bricks have been observed, representing $\sim$20\% of the total expected survey data.  The spatial positions of these Year 1 bricks are shown in Fig.~\ref{footprint}, and we discuss the cluster analysis results from these data in Sec.~\ref{clusterwork}.

In addition to the HST data, there is a remarkable amount of ancillary imaging and spectroscopy of M31 that is already available or will be obtained as part of the PHAT survey.  Imaging datasets from GALEX, Swift/UVOT, Spitzer, and Herschel extend wavelength coverage further into the UV and IR, while observations from CARMA and the EVLA will provide important gas-phase diagnostics.  Finally, red-sensitive spectroscopy of RGB stars from Keck/DEIMOS will improve constraints on galactic kinematics, while MMT/Hectospec observations will provide critical follow-up cluster spectroscopy.

The stellar cluster results will also contribute to other science goals of the PHAT survey, which include placing constraints on the high-mass ($>$ 5 $\rm M_{\odot}$) stellar initial mass function and the calibration of stellar evolution models.  Additionally, the survey goal to derive a spatially-resolved star formation history of the field population of M31 will allow for an interesting comparison between field and cluster age distributions and formation history.

\section{Stellar Cluster Survey\label{clusterwork}}

\begin{figure}
\center
\includegraphics[scale=0.48]{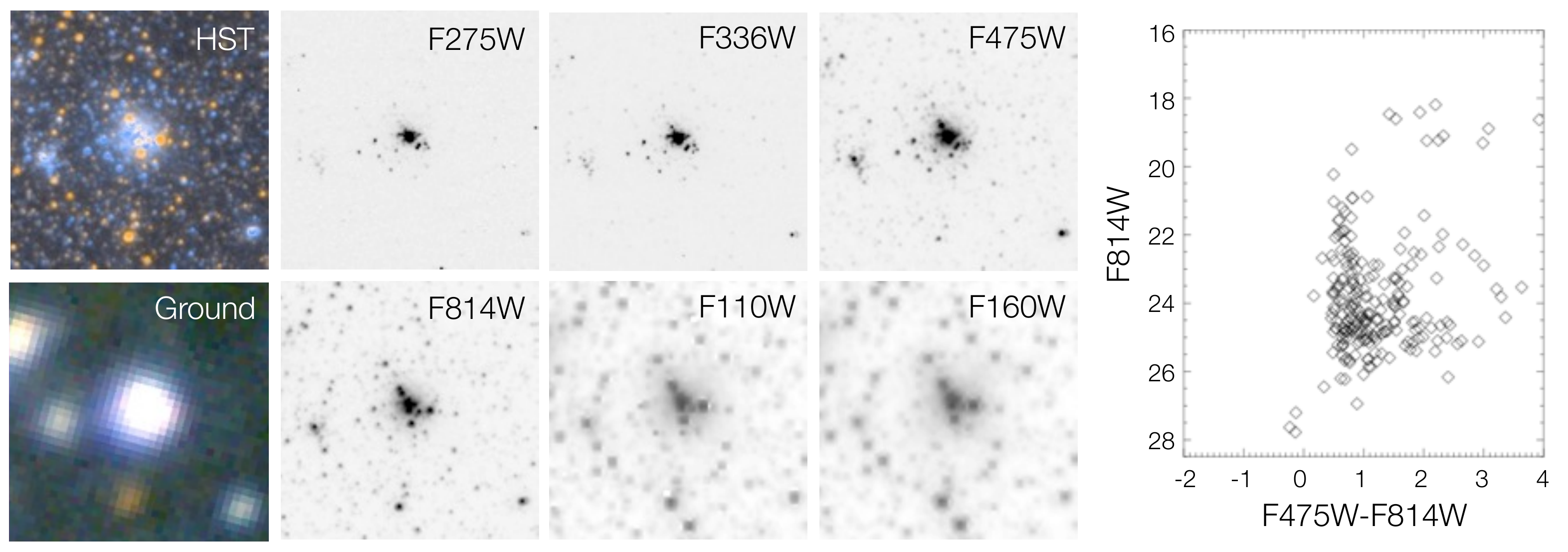} 
\caption{\label{cluster} Image cutouts and optical CMD of cluster B256D showing the data quality available from the PHAT survey imaging.  Image cutouts are 10"$\times$10".  We also include an optical color composite of ground-based imaging from the Local Group Survey \citep{Massey06} for comparison.}
\end{figure}

Our study of M31 stellar clusters is currently underway, with early work focusing on cluster identification and characterization.  For PHAT, we chose to begin the cluster identification process with a systematic by-eye image search, and are currently implementing two automated search techniques to complement the by-eye work.  This approach allows for the careful construction of a well-vetted cluster catalog \citep[following comparable work in M31 by][]{Krienke07, Krienke08, Hodge09, Hodge10}, while simultaneously providing an optimal training set to help refine the automated identification techniques.  Proper calibration of the automated cluster finding routines is required, because unlike most automated extragalactic cluster searches, our clusters are resolved into individual stars (as faint as $\rm M_{F814W} \sim +0$).   For an example of the cluster images provided by PHAT, see Fig.~\ref{cluster}.  Instead of searching for slightly extended point sources, our search routines must account for both cluster components: the resolved stars and the underlying unresolved light.  Our automated identification routines will take advantage of both the resolved and unresolved light components to obtain the best possible identification results.

\begin{figure}
\center
\includegraphics[scale=0.39]{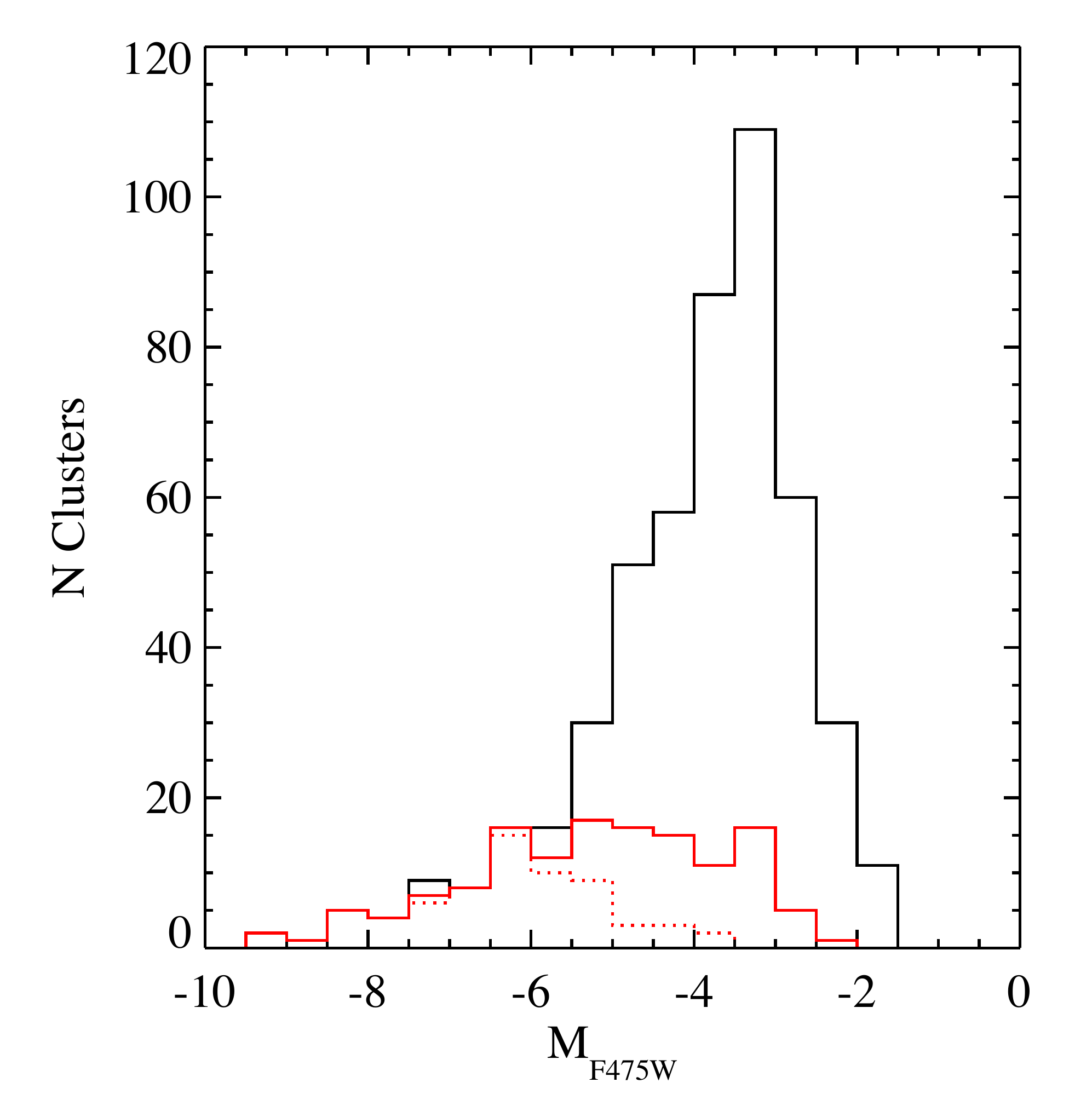} 
\caption{\label{histogram} A histogram showing the F475W absolute magnitude distributions of our PHAT by-eye clusters (black) and previously cataloged clusters (red) that lie within the Year 1 imaging.  The subset of previously known clusters that were discovered using ground-based data are represented by the dotted histogram, while the remaining, fainter known clusters were discovered using limited HST imaging available before PHAT.}
\end{figure}

The first results of our PHAT by-eye search show the great potential of this dataset.  From the Year 1 imaging, we have preliminarily identified $\sim$500 likely clusters, whose spatial distribution is shown in Fig.~\ref{footprint}.  Our current cluster catalog represents a factor of $\sim$3 increase in the number of known clusters for the same area \citep[previously 139 clusters;][and references therein]{Caldwell09}.  As illustrated in Fig.~\ref{histogram}, we find that previous ground-based catalogs were generally complete for bright clusters ($\rm M_{F475W} < -6$), whereas HST imaging allows us to identify clusters more than $\sim$3 mag fainter.  In terms of the faint end of the cluster distribution, we have greatly increased the number of cataloged faint clusters as a result of the marked increase in HST imaging coverage available through PHAT, compared to the limited number of objects discovered through previous, targeted HST observations.  Extrapolating from these results, we estimate the final PHAT stellar cluster sample will include $\sim$2000 clusters, sampling a mass range down to $< $1000 $\rm M_{\odot}$.

Understanding the completeness characteristics of our cluster sample is a high priority considering the population-wide questions we plan to address.  Sample completeness will be easier to quantify once we have incorporated automated cluster-finding routines, but even as part of the by-eye search, we employ artificial cluster tests to make completeness measurements.  Following the philosophy of completeness testing developed for stellar photometry, we insert samples of artificial clusters into the same reduction and analysis pipeline we use for the by-eye search.  This allows us to make an assessment of sample completeness as a function of cluster age, mass, and galactic environment (due to the effects of crowding and extinction).

In addition to understanding sample completeness, our survey has placed great importance on deriving robust age and mass measurements.  One advantage of the PHAT dataset is that it provides us the opportunity to determine cluster characteristics using multiple techniques, and to test for potential systematic differences in the results.  Specifically, we plan to compare age and mass determinations obtained by fitting integrated light measurements to stochastically sampled models \citep[e.g.,][]{Fouesneau10}, through fitting cluster spectroscopy \citep[e.g.,][]{Caldwell09, Caldwell11}, and using color-magnitude diagram (CMD) analysis of individual resolved stars \citep[e.g., ][]{Dolphin02}.  Using this multi-method approach, we will test for consistency between the different techniques, and determine the best methods to derive accurate ages and masses for these clusters, especially in the low to intermediate mass regime.

We expect to publish the first PHAT cluster results in the coming months.  Our Year 1 cluster catalog and accompanying six-band integrated photometry will be included in Johnson et al. (in prep.), while first age and mass assessments will appear in a subsequent paper (Fouesneau et al., in prep.).  After these publications, we plan to periodically expand and update the cluster catalog as additional data arrives over the next three years.  Our team looks forward to reporting the results of our work studying the cluster initial mass function, cluster disruption processes, and the environmental dependencies of cluster formation and destruction, among many other studies made possible by the PHAT survey data.

% Do not delete the next line
\small  % Do not delete
%
%%% Comment the following line if you do not have acknowledgments.
\section*{Acknowledgments}   % Do not delete if you declare acknowledgments
%
%%% ACKNOWLEDGMENTS
Support for this work was provided by NASA through grant number HST-GO-12055 from the Space Telescope Science Institute, which is operated by AURA, Inc., under NASA contract NAS5-26555.

%
% Do not delete the next few lines
%************************************************************************************%
%%%%\begin{thebibliography}{}
%%%%\small
%
%% BIBLIOGRAPHY
%% BIBLIOGRAPHY
%%%%\bibitem{baraffe09}{Baraffe, I., Chabrier, G., Gallardo, J. 2009, ApJ, 702, 27}
%%%%\bibitem{chabrier07}{Chabrier, G., \& Baraffe, I. 2007, ApJ, 661, L81}
%
%
% Do not delete next few lines
%%%%\end{thebibliography}
%************************************************************************************%
%
% Do not delete the next few lines

\bibliographystyle{aa}
\bibliography{mnemonic,ref_Johnson_LC}

\end{document}